\documentclass[twocolumn,aps,prb,superscriptaddress,showpacs]{revtex4}
\usepackage{graphicx}

\begin{document}

\author{Yaroslav Tserkovnyak}
\affiliation{Lyman Laboratory of Physics, Harvard University, Cambridge, Massachusetts 02138}
\author{Arne Brataas}
\affiliation{Department of Physics, Norwegian University of Science and Technology, N-7491 Trondheim, Norway}
\author{Gerrit E. W. Bauer}
\affiliation{Department of NanoScience, Delft University of Technology, 2628 CJ Delft, The Netherlands}
\title{Dynamic stiffness of spin valves}

\begin{abstract}
The dynamics of the magnetic order parameters of ferromagnet/paramagnet/ferromagnet spin valves and isolated ferromagnets may be very different. We investigate the role of the nonequilibrium spin-current exchange between the ferromagnets in the magnetization precession and switching. We find a (low-temperature) critical current bias for a uniform current-induced magnetization excitation in spin valves, which unifies and generalizes previous ideas of Slonczewski and Berger. In the absence of an applied bias, the effect of the spin transfer can be expressed as magnetic--configuration-dependent Gilbert damping.
\end{abstract}

\date{\today}
\maketitle


The giant-magnetoresistance\cite{Baibich:prl88} (GMR) in multilayers of metallic ferromagnet (\textit{F}) and normal-metal (\textit{N}) films controls the current flow by a magnetic--field-induced reorientation of the magnetizations. This effect has found applications in, e.g., magnetic-field detectors as read heads of mass storage devices. Modern magnetic storage media and magnetic random-access memories also consist of \textit{F{\rm/}N} composites with information stored by switchable magnetic configurations. Device performance is measured in terms of bit density as well as speed of reading and writing information. A thorough understanding of the magnetization dynamics in \textit{F{\rm/}N} hybrid structures is therefore desirable.

Magnetization reversal is usually achieved by magnetic fields generated by external electric currents. In small structures, however, much energy is wasted in the form of stray magnetic fields, which motivates consideration of other switching mechanisms. An effect inverse to the GMR (i.e., exerting spin torques on the magnetizations by an electric current passed through the multilayer) appears rather promising for this purpose. The spin torque can lead to magnetization switching at a critical electric current,\cite{Berger:prb96,Sloncz:mmm96} as has been experimentally confirmed.\cite{Myers:sc99,Katine:prl00,Myers:prl02}

Perpendicular spin valves, i.e., \textit{F$_s${\rm/}N{\rm/}F$_h$} trilayer pillar structures with layer thicknesses down to a few monolayers and lateral dimensions in the (sub)micron region, are ideal to study precession and switching phenomena in hybrid systems. When reservoirs are attached on the outer sides, these spin valves can be biased by an electric current perpendicular to the interface planes. \textit{F$_s$} is a \textquotedblleft soft\textquotedblright\ ferromagnetic film with a magnetization that can change easily, whereas \textit{F$_h$} is a \textquotedblleft hard\textquotedblright\ magnetic layer whose magnetization is assumed to be stationary. The relevant variable is then the time-dependent magnetization of the soft layer. In the following, we will consider two scenarios: that of exciting the soft layer by a current bias or driving it by an applied rf magnetic field. In the former case, the layer \textit{F$_h$} can be made stationary by making it much thicker (and therefore more inert for a given spin torque) than \textit{F$_s$} or by using a resistance anisotropy,\cite{Kovalev:prb02} while in the latter case (realized as, e.g., an isolated magnetic bilayer), \textit{F$_h$} can be pinned by an exchange bias or surface magnetic anisotropy.\cite{Urban:prl01} For small enough systems, the magnetic layers are monodomain ferromagnets characterized by two magnetization vectors. Assuming a thick enough spacer \textit{N}, we disregard the static interaction between the ferromagnetic layers, but not the dynamic coupling\cite{Heinrich:prep} induced by the spin pumping.\cite{Tserkovnyak:prl021}

Slonczewski\cite{Sloncz:mmm96} and Berger\cite{Berger:prb96} were the first to predict new time-dependent effects in spin valves. Both authors have realized that a current flowing through a spin valve causes a spin transfer through the nonmagnetic spacer, inducing spin torques on the ferromagnets. In addition, Berger predicted that the two ferromagnets interact via spin transfer even in the absence of an applied electric current, resulting in a significant contribution to the Gilbert damping of the magnetization dynamics. He further demonstrated that a sufficiently large electric current can induce coherent spin-wave emission in the ferromagnet, an idea which was later supported experimentally.\cite{Tsoi:nat00} The condition for spin-wave emission\cite{Berger:prb96} is similar to the criterion for the magnetization switching due to Slonczewski,\cite{Sloncz:mmm96} who treated the Gilbert damping parameter as a phenomenological constant. In Ref.~\onlinecite{Berger:prb96}, a dependence of the damping parameter in spin valves on the relative magnetization angle was found. Some of Berger's and Slonczewski's results as well as the underlying theoretical models were thus not consistent with each other. In this report we offer an alternative theory, which both unifies and extends the seminal work of these pioneers.

Based on the concept of parametric spin pumping, we demonstrated in Ref.~\onlinecite{Tserkovnyak:prl021} that the magnetization motion of a ferromagnetic layer is damped by emitting (pumping) spins into adjacent conductors. The presence of a second ferromagnet can considerably affect the relaxation of the pumped spins, and therefore the magnetization dynamics, as discussed below. We combine adiabatic spin pumping with magnetoelectronic circuit theory\cite{Brataas:prl00,Bauer:prb03} to provide a self-contained framework for spin transfer in spin valves. The main results presented here are the critical current bias for a low-temperature magnetization instability and the configuration-dependent Gilbert damping parameter. In terms of conductance parameters accessible to first-principles calculations\cite{Xia:prb01} and combined with micromagnetic simulations, the full range of the precession and switching dynamics can then be studied in principle. 

\begin{figure}
\includegraphics[width=8cm,clip=]{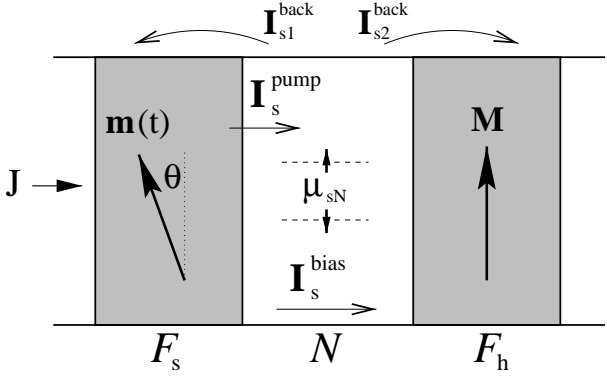}\caption{\label{valve}Schematics of a current-biased spin valve. The symbols are explained in the text}
\end{figure}

We consider the system sketched in Fig.~\ref{valve}. The \textit{F$_s${\rm/}N{\rm/}F$_h$} trilayer is sandwiched between two normal-metal contacts sustaining a charge current bias $J$. The soft layer \textit{F$_s$} magnetization $\mathbf{m}$ will start moving from its equilibrium direction at a critical value $J_c$ (depending on the applied magnetic field). Thermal activation facilitates current-induced magnetization switching,\cite{Myers:prl02} but we focus here on the low-temperature regime. The generalized Landau-Lifshitz-Gilbert equation\cite{Gilbert:pr55,Sloncz:mmm96} for the magnetization direction $\mathbf{m}(t)$ of \textit{F$_s$} in the presence of a spin current $\mathbf{I}_s$ flowing out of \textit{F$_s$} reads
\begin{equation}
\frac{d\mathbf{m}}{dt}=-\gamma\mathbf{m}\times\mathbf{H}_{\text{eff}}+\alpha_0\mathbf{m}\times\frac{d\mathbf{m}}{dt}+\frac{\gamma}{M_sSd}\mathbf{m}\times\mathbf{I}_s\times\mathbf{m} \, ,
\label{llg}
\end{equation}
where $\gamma$ is the absolute value of the gyromagnetic ratio and $\mathbf{H}_{\text{eff}}$ an effective magnetic field which is determined by the applied dc magnetic field and the anisotropy energy (i.e., shape, crystal, and surface anisotropy). \textit{F$_s$} is characterized by $\alpha_0$, its intrinsic (dimensionless) Gilbert damping constant, $M_s$, its saturation magnetization, $d$, its thickness, and $S$, its cross section. The spin current
\begin{equation}
\mathbf{I}_s=\mathbf{I}_s^{\text{exch}}+\mathbf{I}_s^{\text{bias}}
\label{Is}
\end{equation}
consists of the dynamic-exchange current $\mathbf{I}_s^{\text{exch}}$ induced by the spin pumping and of the current $\mathbf{I}_s^{\text{bias}}$ driven by an applied current bias. The former has recently been shown to be responsible for a dynamic coupling between the ferromagnets\cite{Heinrich:prep} and, according to Ref.~\onlinecite{Berger:prb96}, also determines the threshold for spin-wave emission. Alternatively, one can interpret it as a \textquotedblleft dynamic stiffness,\textquotedblright\ which stabilizes the relative magnetization configuration of the spin valve against the torques exerted by $\mathbf{I}_s^{\text{bias}}$ or an applied magnetic field. In high-density metallic systems, the applied voltages and spin accumulations are safely smaller than the Fermi energies, which means that we are in the linear-response regime and both spin currents may be calculated independently of each other. Spin pumping in the outward direction, i.e., into the external connectors, would only increase the intrinsic damping coefficient $\alpha_0$ by a constant value,\cite{Tserkovnyak:prl021} so we disregard it here for simplicity.

Let us first consider the spin current $\mathbf{I}_s^{\text{pump}}$ pumped into the spacer by a time-dependent $\mathbf{m}(t)$ in the absence of an applied bias, $\mathbf{I}_s^{\text{bias}}=0$. The transverse component of the spin accumulation is absorbed in the ferromagnetic layer on the scale of a (transverse) spin-coherence length $\pi/|k_F^\uparrow-k_F^\downarrow|$ in terms of the spin-dependent Fermi wave vectors $k_F^{\uparrow,\downarrow}$ (amounting to only a few monolayers for transition-metal ferromagnets\cite{Brataas:prl00,Stiles:prb02}).  When \textit{F$_s$} is thicker than that,\cite{Tserkovnyak:prl021}
\begin{equation}
\mathbf{I}_s^{\text{pump}}=\frac{\hbar}{4\pi}\left(g_r^{\uparrow\downarrow}\mathbf{m}\times\frac{d\mathbf{m}}{dt}-g_i^{\uparrow\downarrow}\frac{d\mathbf{m}}{dt}\right)\,.
\label{pump}
\end{equation}
Here $g^{\uparrow\downarrow}=g^{\uparrow\downarrow}_r+ig^{\uparrow\downarrow}_i$ is the (dimensionless) mixing conductance of the \textit{F$_s${\rm/}N} interface. For simplicity, the conductance parameters of the two \textit{F{\rm/}N} interfaces are taken to be identical in the following. The mixing conductance is to a good approximation real-valued, i.e., $g_i^{\uparrow\downarrow}\ll g_r^{\uparrow\downarrow}$, at least for transition-metal ferromagnets.\cite{Xia:prb01} $\mathbf{I}_s^{\text{pump}}$ creates a spin accumulation $\text{\protect\boldmath$\mu$}_{sN}$ in \textit{N}, which induces a backflow spin current $\mathbf{I}_{si}^{\text{back}}$ into both ferromagnets $i={1,2}$. According to the circuit theory,\cite{Brataas:prl00,Bauer:prb03} making use of the zero-electric--current condition through the interfaces,
\begin{eqnarray}
\mathbf{I}_{si}^{\text{back}}&=&\frac{1}{4\pi}\left[\frac{2g^{\uparrow\uparrow}g^{\downarrow\downarrow}}{g^{\uparrow\uparrow}+g^{\downarrow\downarrow}}\mathbf{m}_i\left(\Delta\text{\protect\boldmath$\mu$}_{si}\cdot\mathbf{m}_i\right)\right.\nonumber\\
&+&\left.g^{\uparrow\downarrow}\mathbf{m}_i\times\Delta\text{\protect\boldmath$\mu$}_{si}\times\mathbf{m}_i\right]\,.
\label{back}
\end{eqnarray}
Here $g^{ss}$ is the (dimensionless) spin-$s$ interface conductance, $\mathbf{m}_1=\mathbf{m}$, $\mathbf{m}_2=\mathbf{M}$, and $\Delta\text{\protect\boldmath$\mu$}_{si}=\text{\protect\boldmath$\mu$}_{sN}-\mu_{sFi}\mathbf{m}_i$ is the spin-accumulation difference across the \textit{F$_i${\rm/}N} interface. Note that for intermetallic interfaces Sharvin contributions must be subtracted if conductances are computed microscopically by scattering theory.\cite{Bauer:prb03}
The time scale of the magnetization dynamics, $\sim10^{-11}$~s, is much larger than typical electron dwell times in the metallic spacer. Assuming weak spin-flip scattering in \textit{N}, the conservation of spin then implies that
\begin{equation}
\mathbf{I}_s^{\text{pump}}=\mathbf{I}_{s1}^{\text{back}}+\mathbf{I}_{s2}^{\text{back}}\,.
\label{cons}
\end{equation}
The exchange current is given by the difference between the pumped current and the backflow: $\mathbf{I}_s^{\text{exch}}=\mathbf{I}_s^{\text{pump}}-\mathbf{I}_{s1}^{\text{back}}=\mathbf{I}_{s2}^{\text{back}}$.

The longitudinal component of the spin accumulation can penetrate ferromagnets on the scale of the spin-diffusion length $\lambda_{\text{sd}}$. In order to find $\mathbf{I}_{si}^{\text{back}}$, we solve the diffusion equation for the (longitudinal) spin transport in the ferromagnets, assuming that the spin current vanishes on the outer boundaries of \textit{F$_s$} and \textit{F$_h$}. It is shown that the longitudinal spin-accumulation flow into a ferromagnetic slab of thickness $d$ is then governed by an effective conductance $g^\ast$ defined by
\begin{equation}
\frac{1}{g^\ast}=\frac{g^{\uparrow\uparrow}+g^{\downarrow\downarrow}}{2g^{\uparrow\uparrow}g^{\downarrow\downarrow}}+\frac{1}{g_{\text{sd}}\tanh(d/\lambda_{\text{sd}})}
\label{gast}
\end{equation}
in terms of $g_{\text{sd}}=(h/e^2)(S/\lambda_{\text{sd}})(2\sigma^\uparrow\sigma^\downarrow)/(\sigma^\uparrow+\sigma^\downarrow)$,
where $\sigma^s$ is the spin-$s$ conductivity of the ferromagnetic bulk, so that the backflow current, Eq.~(\ref{back}), can be written as
\begin{equation}
\mathbf{I}_{si}^{\text{back}}=\frac{1}{4\pi}\left[g^\ast\mathbf{m}_i\left(\text{\protect\boldmath$\mu$}_{sN}\cdot\mathbf{m}_i\right)+g^{\uparrow\downarrow}\mathbf{m}_i\times\text{\protect\boldmath$\mu$}_{sN}\times\mathbf{m}_i\right]\,.
\label{backN}
\end{equation}
$g^\ast\rightarrow0$ when $d\ll\lambda_{\text{sd}}$, i.e., when the spin-flip relaxation vanishes, or when the ferromagnet is halfmetallic, so that it completely blocks the longitudinal spin flow due to charge conservation. The parameter $\nu=(g^{\uparrow\downarrow}-g^\ast)/(g^{\uparrow\downarrow}+g^\ast)$ characterizes the asymmetry of the absorption of transverse vs longitudinal spin currents. Let us estimate typical values of $\nu$ for sputtered Co/Cu and Py/Cu hybrids at low temperatures, taking $d=5$~nm. The main difference between the two combinations is the spin-diffusion length in the ferromagnets: Co has a relatively long $\lambda_{\text{sd}}\approx60$~nm, while $\lambda_{\text{sd}}\approx5$~nm is very short in Py.\cite{Bass:mmm99} Using known values for spin-dependent conductivities,\cite{Bass:mmm99} we thus find $g_{\text{sd}}S^{-1}\approx2.7$~nm$^{-2}$ for Co and 16~nm$^{-2}$ for Py. $2g^{\uparrow\uparrow}g^{\downarrow\downarrow}/(g^{\uparrow\uparrow}+g^{\downarrow\downarrow})S^{-1}\approx20$~nm$^{-2}$ for Co/Cu interfaces\cite{Xia:prb01} and we expect the value for Py/Cu to be similar. Finally, taking $g^{\uparrow\downarrow}S^{-1}\approx28$~nm$^{-2}$ for the Co/Cu interface\cite{Xia:prb01} and 15~nm$^{-2}$ for Py/Cu,\cite{Bauer:prb03} we find $\nu\approx0.98$ for Co/Cu and $\nu\approx0.33$ for Py/Cu.

With $\nu$ the same for both layers, we solve for the spin accumulation in the normal metal, $\text{\protect\boldmath$\mu$}_{sN}$, in the absence of applied current, $J=0$, by using the spin conservation, Eq.~(\ref{cons}), and Eq.~(\ref{backN}) for the backflow in terms of $\text{\protect\boldmath$\mu$}_{sN}$. We then finally arrive at
\begin{equation}
\mathbf{I}_s^{\text{exch}}=\frac{1}{2}\left[\mathbf{I}_s^{\text{pump}}-\nu\left(\mathbf{I}_s^{\text{pump}}\cdot\mathbf{M}\right)\frac{\mathbf{M}-\nu\mathbf{m}\cos\theta}{1-\nu^2\cos^2\theta}\right]\,.
\label{exch}
\end{equation}
Semiclassically, this equation can be understood as a multiple scattering of spin current between the interfaces at which the longitudinal part is reflected with probability $P\propto1+\nu$ and the transverse component with $P\propto1-\nu$. We have taken the spacer to be ballistic, so that $\text{\protect\boldmath$\mu$}_{sN}$ is uniform. Otherwise, the exchange current will be suppressed by diffuse scattering in the interlayer \textit{N}. It is straightforward to extend our theory to take this into account by, e.g., solving the spin-diffusion equation in the spacer and using the same boundary conditions, Eqs.~(\ref{pump}) and (\ref{backN}), as above.

\begin{figure}
\includegraphics[angle=-90,width=8cm,clip=]{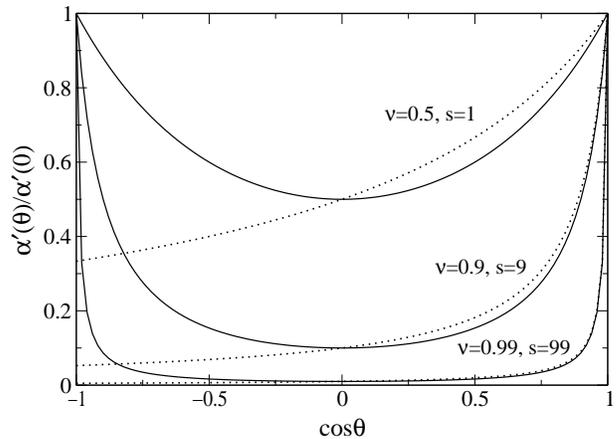}\caption{\label{cone}Solid lines are our prediction for the precession-cone angle dependence of the Gilbert damping parameter [Eq.~(\ref{or})] and the dotted lines are Berger's [Eq.~(\ref{B})]. The lower (solid) line is representative for Co, while the upper for Py, assuming thickness of 5~nm. We expect Fe and Ni to be characterized by the two lower (solid) lines.}
\end{figure}

The magnetization dynamics (in the absence of an applied bias) is determined by substituting $\mathbf{I}_s^{\text{exch}}$ into the LLG equation, which thus has a damping term that cannot be modeled by a constant effective Gilbert parameter. We now analyze the configuration dependence of the damping in more detail, which is experimentally accessible by the FMR line-width broadening at high rf intensities\cite{Heinrich:priv} (and therefore finite \textquotedblleft precession cones\textquotedblright). For $\mathbf{m}$ precessing around $\mathbf{M}$,
\begin{equation}
\mathbf{m}\times\mathbf{I}^{\text{exch}}_s\times\mathbf{m}=\frac{g^{\uparrow\downarrow}}{8\pi}\left(1-\nu\frac{\sin^2\theta}{1-\nu^2\cos^2\theta}\right)\mathbf{m}\times\frac{d\mathbf{m}}{dt}\,.
\label{pr}
\end{equation}
The angular dependence of the additional Gilbert damping parameter due to the exchange spin current then reads
\begin{equation}
\alpha^\prime(\theta)/\alpha^\prime(0)=1-\nu\sin^2\theta/(1-\nu^2\cos^2\theta)\,,
\label{or}
\end{equation}
where $\alpha^\prime(0)=\gamma\hbar g^{\uparrow\downarrow}_r/(8\pi M_sSd)$ is the damping enhancement in a collinear configuration.
Interestingly, this result bares similarity with Berger's\cite{Berger:prb96}
\begin{equation}
\alpha^\prime(\theta)/\alpha^\prime(0)=1/[1+s(1-\cos\theta)]\,,
\label{B}
\end{equation}
where $s\propto\tau_{\text{sf}}$, a characteristic spin-flip time, and his form of $\alpha^\prime(0)$ is similar as well.\cite{Berger:prb96} Expressions (\ref{or}) and (\ref{B}) are compared in Fig.~\ref{cone}: they have the same small-angle asymptotics, but are qualitatively different at large angles (in particular, the latter is symmetric with respect to $\cos\theta$ while the former is not). At small $\theta$, we can rewrite Eq.~(\ref{or}) to exactly reproduce Eq.~(\ref{B}) after identifying $s=\nu/(1-\nu)$. As mentioned above, $\nu$ is close to 0.98 for cobalt and $s$ should be of the order of 100 ($s=333$ is found in Ref.~\onlinecite{Berger:prb96} for Co/Cu with Co 1.5~nm thick, which remarkably would be quite similar to our estimate for this thickness), so that the lower solid line in Fig.~\ref{cone} represents the damping for Co. The \textquotedblleft precessional stiffness\textquotedblright\ is thus significantly reduced for angles which only slightly deviate from the collinear configurations (we expect this conclusion to be also true for Fe and Ni). Modeling of the magnetization dynamics with a constant damping parameter is thus not allowed for sufficiently thin magnetic layers. For permalloy, on the other hand, the precessional stiffness is expected to remain significant for all angles, see the upper solid line in Fig.~\ref{cone}. This implies that the magnetization reversal has higher energy-dissipation power, but can occur faster than in cobalt in field-induced switching. If $\mathbf{m}$ moves away from $\mathbf{M}$, i.e., only the relative angle $\theta$ changes, then $\mathbf{I}_s^{\text{pump}}\perp\mathbf{M}$ and Eq.~(\ref{exch}) reduces to $\mathbf{I}_s^{\text{exch}}=\mathbf{I}_s^{\text{pump}}/2$. The \textquotedblleft tilting stiffness\textquotedblright\ has thus an angle-independent enhancement with respect to the intrinsic Gilbert damping, which is given by the same expression as $\alpha^\prime(0)$, i.e., the damping in a collinear configuration.

Introducing an applied current bias gives an additional control over the magnetization dynamics. When the conductance parameters of the spin valve are symmetric, the bias-induced spin transfer $\mathbf{I}_s^{\text{bias}}$ is coplanar with the magnetization directions, $\mathbf{I}_s^{\text{bias}}=I_s^{\text{bias}}(\mathbf{m}+\mathbf{M})/(2\cos\theta/2)$, $\theta$ being the angle between $\mathbf{m}$ and $\mathbf{M}$. This is clear after expanding the spin current as $\mathbf{I}_s^{\text{bias}}(\mathbf{m},\mathbf{M})=f_{11}(\cos\theta)\mathbf{m}+f_{22}(\cos\theta)\mathbf{M}+f_{12}(\cos\theta)\mathbf{m}\times\mathbf{M}$ and noting that $\mathbf{I}_s^{\text{bias}}(\mathbf{m},\mathbf{M})=\mathbf{I}_s^{\text{bias}}(\mathbf{M},\mathbf{m})$, which implies $f_{11}\equiv f_{22}$ and $f_{12}\equiv0$. The electric current corresponding to a given spin-current bias depends on $\theta$ and can be calculated readily by circuit theory.\cite{Brataas:prl00,Bauer:prb03}

Eqs.~(\ref{llg}), (\ref{Is}), (\ref{exch}) and the form of the bias current completely determine the dynamics of $\mathbf{m}(t)$. The exchange induced by the spin pumping causes relaxation toward an equilibrium configuration, while the bias current can either relax or excite a perturbation from an equilibrium, depending on the sign of $J$.
In the process of, e.g., switching, the trajectory of $\mathbf{m}(t)$ can become very complicated. While in this report we outline the general formalism, a detailed numerical study of the magnetization dynamics will be carried out elsewhere. In the remainder of this paper we discuss the critical current bias at which a collinear equilibrium configuration becomes unstable.

Near a collinear configuration, Eq.~(\ref{exch}) simplifies to $\mathbf{I}_s^{\text{exch}}=\mathbf{I}_s^{\text{pump}}/2$. Let $\mathbf{m}$ (circularly) precesses around $\mathbf{M}$ with the FMR frequency $\omega$: $\mathbf{m}\times d\mathbf{m}/dt=\omega\mathbf{m}\times\mathbf{M}\times\mathbf{m}$. The total (projected) spin current in the Gilbert form then reads:
\begin{equation}
\mathbf{m}\times\mathbf{I}_s\times\mathbf{m}=\left[\hbar g^{\uparrow\downarrow}/(8\pi)+I_s^{\text{bias}}/(2\omega)\right]\mathbf{m}\times\frac{d\mathbf{m}}{dt}\,.
\end{equation}
An instability is reached when the effective Gilbert damping coefficient becomes negative. The critical bias is thus given by
\begin{equation}
I_{s,c}^{\text{bias}}=\left[g^{\uparrow\downarrow}/(4\pi)+2\alpha_0M_sSd/(\hbar\gamma)\right]\hbar\omega\,.
\end{equation}
Neglecting the first term in the brackets of the above expression, we obtain result analogous to Slonczewski's,\cite{Sloncz:mmm96} while neglecting the second term, we get a condition similar to Berger's spin-wave emission criterion.\cite{Berger:prb96} The spin-pumping contribution (first term) is comparable with the intrinsic damping (second term) for films with thickness $d$ of several nanometers,\cite{Tserkovnyak:prl021,Urban:prl01} with the former dominating for very thin films.

In summary, we have developed a general theoretical framework for the low-temperature magnetization dynamics in small spin valves, unifying and extending pioneering work by Slonczewski\cite{Sloncz:mmm96} and Berger.\cite{Berger:prb96} The non-equilibrium spin torque induced by the bias current and the enhanced Gilbert constant due to the spin pumping must be treated on equal footing. When the memory magnetic element is sufficiently thin ($d<10$~nm), the nontrivial dependence of the damping on both the static and dynamic configurations of the system can importantly modify the magnetization dynamics. We derived the dependence of the Gilbert damping of \textit{F$_s$} on the precession-cone angle, which can also be measured by the FMR.\cite{Heinrich:priv} Micromagnetic simulation codes should take these effects into account as the device and magnetic bit dimensions decrease down to the nanometer scale.

We would like to thank B.~I.~Halperin and B.~Heinrich for many
stimulating discussions. This work was supported in part by the NEDO
International Joint Research Grant Program \textquotedblleft
Nano-magnetoelectronics,\textquotedblright\ NSF Grant DMR 02-33773, and
the FOM.

\end{document}